\documentclass[conference,10pt]{IEEEtran}%
\usepackage{amsmath}
\usepackage{epsfig}
\usepackage{amssymb}
\usepackage{paralist}
\usepackage{url}
\usepackage{balance}
\balance
\usepackage{color}
\usepackage[noend]{algorithmic}
\usepackage[ruled,vlined,boxed,commentsnumbered]{algorithm2e}
 \SetAlCapSkip{1.0em}
\DeclareMathOperator{\SINR}{SINR}
\newtheorem{boldtheorem}{\textbf{Theorem}}
\newtheorem{lemma}{{Lemma}}
\def\sources{{\cal S}}
\def\actree{{\textbf T}} 
\def\processing{{\chi}}
\def\queryset{{\cal Q}}
\def\period{{\textbf p}}
\def\delay{{\textbf d}}
\def\deadline{{\textbf f}}
\def\release{{\textbf a}}
\def\initiaload{{\ell}}
\def\weight{\textbf{w}}
\def\ie{\textit{i.e.}\xspace}
\def\etal{\textit{et al.}\xspace}
\def\eg{\textit{e.g.}\xspace}
\def\maxlink{\textbf r}
\def\load{{{{\cal L}}}}
\def\pathloss{{\kappa}}

\def\regionlenth{{\lambda}}
\def\qnum{{c}} 

\author{\IEEEauthorblockN{Xiaohua Xu}
\IEEEauthorblockA{EECS Department\\
The University of Toledo\\
Toledo, OH 43606\\
Email: Xiaohua.Xu@utoledo.edu}
\and
\IEEEauthorblockN{Xiang-Yang Li}
\IEEEauthorblockA{
Department of Computer Science\\
Illinois Institute of Technology\\
Chicago, IL 60616\\
Email: xli@cs.iit.edu}
\and
\IEEEauthorblockN{Min Song}
\IEEEauthorblockA{EECS Department\\
The University of Toledo\\
Toledo, OH 43606\\
Email: Min.Song@utoledo.edu}}
\begin{document}

\title{Real-time Data Collection Scheduling in Multi-hop Wireless Sensor Networks}
\maketitle

\begin{abstract}
We study real time periodic query scheduling for data collection in multihop
 Wireless Sensor Networks (WSNs).
Given a set of heterogenous data collection queries in WSNs,
 each query requires
 the data from the source sensor nodes to be collected to the control center within a certain end-to-end delay.
We first propose almost-tight necessary conditions for a set of
 different queries to be schedulable by a WSN.
We then develop a family of
 efficient and effective data collection algorithms that can
 meet the real-time requirement 
  under resource constraints by addressing
 three tightly coupled tasks:
 (1)  routing tree construction for data collection,
 (2) link activity scheduling, and (3) packet-level scheduling.
Our theoretical analysis for the schedulability
 of these algorithms show that they can achieve a constant fraction of
 the maximum schedulable load.
For the case of overloaded networks
 where not all queries can be possibly satisfied, we
 propose an efficient approximation algorithm to select queries to
   maximize the total weight of selected schedulable queries.
The simulations
 corroborate our theoretical analysis.

\end{abstract}


\section{Introduction}
\label{sec:introduction}
Recent years have seen the emergence of
wireless sensor networks (WSNs).
WSNs are
 deployed to monitor various aspects of the environment,
 such as temperature and light.
The WSNs are also being deployed in a wide variety of other
applications. For WSN applications,
 the data in the sensors are often streamed to
 a control center (called sink).
This process is called data collection.
For most control applications, the observed
 events and consequently the data from the source sensors must be
 collected at the control center within a certain
 delay.
A key challenge then in WSNs is to meet the
 end-to-end delay requirement of control applications under wireless
 interferences and the severely limited resource constraints of WSNs.

Multitudes of protocols have been proposed in the literature for data
 collection in WSNs that balance the communication cost, delay, and reliability
\cite{lee2008teo}.
 However, not much effort has been paid into the design of real-time data collection schemes
 that provide end-to-end performance guarantees for periodic queries.
In this paper, we concentrate on designing effective
 scheduling of activities of nodes to satisfy
 multiple heterogeneous queries.
Given a set of sensor nodes
 and a sink node,
 the sink node issues a set of periodic queries, each
 has a period, initial release time
 and relative deadline requirement for collecting the corresponding data.
The sink node expects to receive the corresponding data from all sensor nodes in time.
Given an arbitrary interference model,
 the objective is to jointly design routing
 and an
 interference-aware schedule of activities for all nodes (\ie, when to
 transmit and what packets to transmit) such that the deadlines of all queries are met.

Our main contributions are the
 schedulability test and effective scheduling algorithms
  summarized as follows. 
First, we propose a necessary condition for a set of queries to be
schedulable:
 Theorem~\ref{the:c_x} summarizes a necessary condition for data
collection queries under various interference models.

Second, 
 we design efficient algorithms for constructing a routing tree for
each of queries, scheduling node activities for each wireless node,
and packet scheduling. We theoretically prove that the schedulable
queries by our methods achieve a load that is within a constant
factor of the maximum schedulable load. Based on the proposed
algorithms, in Theorem~\ref{the:c_3}, we present a sufficient condition for
 schedulability of data collection queries under various interference models in WSNs.

Third,
  we propose an efficient query-selection algorithm by carefully selecting
 a subset of queries such that
 the total weight of selected  queries (that are schedulable by our
 algorithms) is at least a constant fraction of the optimum solution
 when the load of all queries exceed the network capacity (\ie,
 the WSN is overloaded with queries from control applications).

Finally, we conduct extensive
 simulations to validate proposed algorithms. Our simulation
 results
 in TinyOS
 corroborate our theoretical analysis.

\textbf{Related Work:}
\label{sec:review}
Real-time scheduling (see \cite{liu2000real} and references
therein) has been extensively studied in the literature.
The two most representative classes of scheduling algorithms are
rate-monotonic (RM) scheduling and Earliest Deadline First (EDF) scheduling.
RM algorithms assign static-priorities to jobs on
 the basis of the cycle duration of the jobs.
In the pioneering work \cite{liu-layland-jacm73},
 Liu and Layland  proposed a RM
 algorithm in a single processor,
 and the first sufficient condition for schedulability of a set of queries.
This result has been further extended in
\cite{leung1982complexity,shih1993modified}.
On the other hand, EDF is a dynamic scheduling algorithm. EDF and
its several extensions \cite{sivaraman2000traffic,sivaraman2000providing,sivaraman2001end}
 have been proposed 
 to guarantee the end-to-end delay of packets.

Only a few work
 have studied  the ``real-time''
 group communication scheduling in multi-hop WSNs.
Chipara \etal\cite{chipara2006dynamic, chipara2007real} studied the
 real time query scheduling for data aggregation by assuming a pre-given routing
 tree. However, the methods do not provide a theoretical performance
 assurance.
Xu \etal \cite{xu2012efficient} studied periodic query scheduling for
data aggregation with minimum delay under various wireless interference models.

The problem of sporadic query scheduling in the network for data
 collection under various interference models has
 been extensively studied recently
 \cite{wang2010tosn,annamalai2003tbc,gandham2006dmt,kesselman2006fda,upadhyayula2003lla}.
One-shot query scheduling for data processing with
 minimum delay has been proven
 to be NP-hard \cite{annamalai2003tbc,gandham2006dmt,kesselman2006fda}.
A collision-free scheduling method for data collection is proposed
 in \cite{lee2008teo}, which aims at optimizing energy consumption and reliability.

The rest of the paper is organized as follows. Section
\ref{sec:model} presents the system model.
Section \ref{sec:col}
 presents schedulability results on data collection queries under various interference models.
Section \ref{sec:weighted} studies the query scheduling in
overloaded networks. We present our simulation results in
 Section \ref{sec:simulation} and conclude the paper in Section
\ref{sec:conclusion}.

\section{System models}
\label{sec:model}
Consider a WSN as a graph $G=(V,E)$, consisting of a set $V$ of $n$
 sensor nodes
 where $v_s\in V$ is the sink node and $E$ is the set of
 communication links.
Two nodes can communicate with each other {if}
 they are within the transmission range of each other.
A set of links can transmit simultaneously if and only if they are
 interference free.
 Several interference models such as Protocol Interference Model (PrIM),  RTS/CTS Model, and Physical
 Interference Model (PhIM) or the Signal-to-Interference-plus-Noise
 Ratio model (SINR model) have been considered in the literature and used in transmission
scheduling studies.
In PrIM \cite{gupta2000capacity}, each node
 $v_i$, in addition to have a uniform transmission range (scaling to $1$),
 has an \textit{interference range} $\rho$ such that
 any node $v_j$ will be interfered by the signal from
 $v_i$ if $\|v_i-v_j\| \le \rho$ and node $v_j$ is \emph{not}
 the intended receiver of the transmission from $v_i$.
In the RTS/CTS  model~\cite{alicherry2005joint}, for every
 pair of transmitter and receiver,
 all nodes that are within the interference range of either the transmitter
 or the receiver cannot transmit.
In PhIM \cite{goussevskaia2007complexity}, there is a threshold
value $\beta>0$,
 such that a node $v_j$ can correctly
 receive the data from a sender $v_i$ if and only if
 the signal to interference plus noise ratio at the receiver satisfies
 \[\SINR (v_i, v_j) =\frac{P_i\cdot d^{\pathloss}_{i,j}}{ N_0+ \sum_{k\in I}P_k\cdot d^{-\pathloss}_{k,j}}\ge\beta.\]
Here
 $d_{k,j}$ is the Euclidean distance $\|v_k-v_j\|$, $N_{0}> 0$ is the
 background noise, $P_i$ is the
 transmission power of node $i$ (we assume the transmission power is a
 constant, \ie, $P_i=P$), $I$
 is the set of actively transmitting nodes when node $v_i$ is
 transmitting, and $\pathloss > 2$ is the path loss exponent.

Assume the control application issues a set
 of heterogenous  data collection queries, and
 source nodes generate source data periodically at specified
 data rates.
 In practice, queries could be different in many aspects.
The $i$-th query can be characterized as follows: let $\sources_i
\subseteq V$ denote a subset of source nodes, each source node generates data to
answer this query. We assume that each source node $v\in\sources_i$
 will generate a data unit to be
 collected to the sink $v_s$ periodically.
We assume that it takes $\processing_i$ time to transmit a data unit
for the $i$-th query over any link in the network. Here
$\processing_i$ could be different for different queries. For
simplicity, we assume that $\processing_i$ already takes into
account the link reliability, data preparing time at nodes,  and
data size variety for answering queries.

The $i$-th query  will be initially released at time
 $\release_i$ and will have an end-to-end delay
 requirement  $\delay_i$ for receiving the answer.
In other words, the sink should receive the corresponding data before time
 $\deadline_i=\release_i + \delay_i$.
We  assume that the $i$-th query has a period $\period_i$;
 then, the $t$-th instance of this query   will be released
 at time $\release_i + (t-1) \cdot \period_i$ and the deadline for receiving
 the data for this instance is $\deadline_i^t=\release_i + (t-1) \cdot \period_i + \delay_i$.

Two different questions will be answered in this work. First, given
a set of $\qnum$ queries $\queryset$ for data collection,
 each with its own period $\period_i$,
processing time  $\processing_i$,  end-to-end deadline
$\deadline_i$, and a set of sources nodes $\sources_i \subset V$,
 whether the set of queries can be satisfied, and if so, design
 effective routing and scheduling algorithms to meet the specified requirements.
The second type of questions is to
design routing and scheduling
 protocols that will maximize the total weight of scheduled queries
 when we cannot schedule all queries successfully and each query is
 associated with a positive weight.


\section{Real-time Schedule for Data Collections}
\label{sec:col}

We first propose both necessary conditions and
 sufficient conditions for schedulability of a given set of data collection queries.
We then develop efficient  routing protocols, link scheduling, and
packet
 scheduling methods to satisfy a schedulable set of queries.

\subsection{Necessary Conditions for Schedulability}
Our study of necessary conditions and later sufficient conditions
 for schedulability rely on
 the concepts of \emph{initial load} and \emph{relay
 load} of a node (and/or a region) \cite{xu2012efficient}.
Let us first review the concept of \emph{initial load}. Given a
WSN  $G=(V,E)$ and a set of queries $\queryset$, the initial load of
a node
 $u\in V$ is defined as
 $\initiaload_{G,  \queryset}(u) = \sum_{u\in \sources_j} \frac{ \processing_j}{\period_j}$,
where $\processing_j$ is the processing time, $\period_j$ is the
 period, and $\sources_j\subseteq V$
 is the set of source nodes of the $j$-th query.
If we denote \emph{region} as any continuous area in a
two-dimensional plane, the initial load of a region $g$ is defined
as the summation of the
 initial loads of
 all nodes in this region $g$, \ie
 $\initiaload_{G,  \queryset}(g_{v,h})=\sum_{u\in V(g)}
 \initiaload_{G,  \queryset}(u)$ where
$V(g)$ consists of all nodes from $V$ lying in the region $g$.

We will focus on the initial load of a special
region
 (called \emph{interference-aware region})
 which is a square in a two-dimensional plane, with the \emph{interference-aware radius} as its side-length.
Given an interference model $\cal M$, the {interference-aware
radius}
 $\regionlenth({\cal M})$ is the
 maximum possible distance between two senders such that the
 corresponding two links will interfere with each other under $\cal M$.
This means that a set of nodes can transmit concurrently without
interference  if the distance between
 any pair of nodes is greater than
$\regionlenth(\cal M)$.
We can compute $\regionlenth(\cal M)$ based
on the parameters of the
 model $\cal M$,
We then partition the two-dimensional plane by using a set of
vertical lines
 $a_i: x=i \cdot \regionlenth({\cal M})$ where $i\in \mathbb{Z}$
 and horizontal lines $b_j: y=j\cdot \regionlenth({\cal M})$ where
 $i\in \mathbb{Z}$.
Here $\mathbb{Z}$ represents the set of all integers
 and $i, j\in \mathbb{Z}$ is called the index of vertical line $a_v$ and
 horizontal line $b_h$.
We denote the {interference-aware region} formed by a pair of neighboring vertical lines
$a_{i}, a_{i+1}$ and a pair of neighboring horizontal lines $b_{j},
 b_{j+1}$ as $g_{i, j}$.

To schedule the nodes' transmissions,
 for a clique in the node-conflict graph where any pair of nodes cannot transmit concurrently,
 the summation of nodes' initial loads in the clique can not exceed one.
Generally, for any interference-aware region
  where the maximum number of nodes in that region that can
  transmit concurrently is a constant $c_1({\cal M})$,
 the initial load of this region is at most $c_1({\cal M})$.

On the other hand,
for the $i$-th query,
 no matter what data collection routing tree is used,
 the sink node needs to receive all the raw data from $\sources_j$.
Thus, the initial load of sink node
 coming from the $i$-th query is exactly $\frac{|\sources_i| \cdot \processing_i}{\period_i}$.
If a set of queries $\queryset$ can be satisfied,
 the initial load of the sink node $\sum_{i} \frac{|\sources_i| \cdot
\processing_i}{\period_i}$ is at most one. Therefore,
 we propose a necessary condition for schedulability as follows.
\begin{boldtheorem}\label{the:c_x}
If a set of data collection queries $\queryset$ under an
interference model ${\cal M}$ is schedulable, then
\begin{equation}
\begin{cases}
\initiaload_{G,  \queryset}(g_{v,h}) &\le c_1({\cal M}), \ \forall g_{v,h}\\
\sum_{i} \frac{|\sources_i| \cdot \processing_i}{\period_i}  & \le 1
\end{cases}
\end{equation}
Here $\initiaload_{G,  \queryset}(g_{v,h})$ is the initial load of
an
 interference-aware region $g_{v, h}$.
Constant $c_1({\cal M}) \ge 1$ is the maximum number of nodes
 that can transmit concurrently
 in any interference-aware region
 under the interference model ${\cal M}$.
\end{boldtheorem}

Henceforth all the proofs will be available in the
technical report \cite{xu-technical-globecom} due to the page limit.

Next, we derive the value of $c_1({\cal M})$ under various
interference models.
Note that for physical interference model,
 the interference-aware radius
$\regionlenth({\cal M})$ is the same as the \emph{maximum transmission
radius} $\maxlink = \sqrt[\pathloss]{\frac{P}{N_0\beta}}$.
The {maximum transmission
radius $\maxlink$} can be perceived as a threshold for communication distances:
a pair of nodes can possibly communicate and thus be
connected \emph{iff} their mutual distance is smaller than the threshold $\maxlink$.
In other words, a node $u$ cannot transmit data to another node $v$
 which is more than $\maxlink$ distance away
 even in the absence of other concurrent transmissions.

\begin{lemma}
\label{lem:c_1_pro}
The constant $c_1({\cal M})$ is given as:
\begin{equation*}
c_1({\cal M})=
\begin{cases}
\frac{16\cdot \rho^2}{(\rho-1)^2} & \text{under PrIM}\\
36   &    \text{under RTS/CTS}\\
\lfloor\frac{2^\pathloss \cdot P}{N_0 \beta^2}\rfloor &  \text{under PhIM}
\end{cases}
\end{equation*}
\end{lemma}

\subsection{Efficient Algorithms for Scheduling Queries}\label{sec:col_alg}
In this section, we design effective algorithms for scheduling data
collection queries under various interference models.
For each data collection query, each node needs to transmit its raw data (if it has) and relay all
received data towards the sink node periodically.

%

The first phase is construct routing trees.
The constructions of routing trees
 are similar under various interference models.
Given a communication graph $G=(V,E)$, we select
 a CDS $\actree_{CDS}$ of $G$ by using an existing approach
 \cite{wan2002infocom}.
We then construct a \emph{spanning tree} $\actree_G$ by connecting
 each node not in the CDS to a neighboring dominator in the CDS.
For the $i$-th query, we construct the routing tree $\actree_i$ based on $\actree_G$ by
 pruning every node $u\in V$ and the corresponding link
 $\overrightarrow{up(u)}$ (the link from $u$ to its parent $p(u)$)
 if the intersection between $\sources_i$ and the subtree of $\actree_{G}$ rooted
 at $u$ (noted as $\actree^u_G$) is empty: $\sources_i\cap
 \actree^u_G=\emptyset$.

Under PhIM,
 we construct routing trees
 in a reduced communication graph instead
of in the original graph.
The definition of reduced communication graph is available in \cite{xu2012TMC}.

The second phase is to construct a real-time \emph{transmission
 plan} for each node after we construct a routing tree for each query.
Observe that a node $u$ is involved in the $j$-th query
 if: (1) $u$ is a source node for this query, \ie, $u\in \sources_i$,
 or (2) $u$ is a relay node for this query.
In either case,  $u\in \actree_i$.
Thus, we test $u\in
\actree_i$ to determine
 whether a node $u$ is involved in the $i$-th query or not.
If $u\in \actree_i$ is true, during each period
$\period_i$,
 node $u$ needs to add a data unit for this query to its transmission plan.
The added packets are either original  or
 relayed packets.
For each node, we store the transmission plan to its buffer.

The third phase is to schedule (or assign) concrete time to each
node
 for transmission,
 and to avoid interference at the same time.
This phase consists of two steps: (1) determine which
region to select nodes from, called an \emph{active region}; (2)
determine which node in an active region to transmit.

First, we color all interference-aware regions
 such that any pair of neighboring regions with the same
 color are separated by $K({\cal M})-1$ regions, where $K({\cal M})$ is a constant
 depending on the interference model.
Clearly, the chromatic number for this coloring method is $c_2({\cal
 M})=K({\cal M})^2$.
To avoid interference, each time we only allow regions with the same
 color to be active.
Specifically, we have $c_2({\cal M}) =4$ under
  PrIM and the RTS/CTS model and $c_2(\cal
 M)$ is a constant under PhIM \cite{xu2012TMC}.
With the help of region coloring, we ensure that if
 only one node is selected from each interference-aware region with
 the same color to transmit, we can avoid interference, irrespective of the positions of
 the receivers.

Second, we assign transmission time to nodes in an active region.
Clearly, a node with more relay load needs to be assigned with more
time. We propose a {linear time assignment} scheme in which each
node in an active
 region is assigned with transmission time proportional to its relay load.
 The linear time assignment scheme is describe as follows.
Given a set of queries $\queryset$ and the corresponding routing trees, we define the {relay load} of a node $u$
as $\load_{G,  \queryset}(u)=\sum_{ \actree_j \ni u} \frac{
 \processing_j}{\period_j}$.
We then define the relay load of a region $g$ as
 the summation of
 all nodes' relay loads in this region:  $\load_{G,  \queryset}(g_{v,h})=\sum_{u\in V(g)} \load_{G,
 \queryset}(u)$, where $V(g)\subseteq V$ is the set of all nodes from $V$ lying
 in region $g$.
The relay load contains both the initial load and the data load
coming from routing.
Thus, the relay load of a node can be perceived
 as the fraction of time for a node
 to be actively transmitting data.
Given a time duration $T$ (here $T>\period_j, \forall j$) when
an interference-aware region $g$ is active,
 we  assign each node $u$ in region $g$ with transmission time $T\cdot
 \frac{\load_{G,  \queryset}(u)}{\load_{G,  \queryset}(g_{v,h})}$.

The details are shown in Algorithm~\ref{alg:interference-aware}
which is performed for every $c_2({\cal M})\cdot T$ time duration.
Then each region is active for exactly $T$ time duration. When a
region is active, we apply linear time assignment to each node in
this region.

{\small
\begin{algorithm}[t]
\label{alg:interference-aware} \LinesNumbered
\caption{Interference-aware node scheduling}
\SetKwInOut{Input}{Input}\SetKwInOut{Output}{Output}
\Input{Routing trees for all queries}

{ $K({\cal M}) \leftarrow \lceil {\sqrt{c_2({\cal M})} }\rceil$\;

\For{each interference-aware region $g_{v,h}$ where $v,h
\in\mathbb{Z}$ and $g_{v,h}$ contains nodes} {
 Assign the region with color:\\
 $\left( v\mod K({\cal M}) \right)\cdot K({\cal M})+ h\mod K({\cal M})$;}

\For{$i=1, \cdots, K({\cal M})$ and $j=1, \cdots, K({\cal M})$}{

 \For{each region $g_{v,h}$ of
 the $i\cdot K({\cal M})+j$-th color where $v,h \in\mathbb{Z}$,
 and $g_{v,h}$ contains nodes}{
\For{each node $u$ in region $g_{v,h}$}{
 assign the node with transmission time:
 $T\cdot \frac{\load_{G,  \queryset}(u)}{\load_{G,  \queryset}(g_{v,h})}$\; }}
} \Return{a set of transmission time for each node.} }
\end{algorithm}
}

The fourth phase is
  to select packet(s) from the node's transmission plan to
  transmit when it is a node's transmission time.
We use a \emph{rate monotonic} \cite{liu-layland-jacm73,
shih1993modified} method to select packets from the node's
transmission plan.

$1)$ All packets of current period have lower priorities than that
of all previous periods.

$2)$ The priorities of all packets of any queries are assigned on a
rate-monotonic basis. In other words, a packet of current instance
for a query with a
 shorter period has a higher priority over
 the packet of current instance for a query with a longer period
 (at absolute time $t$,
 a packet is at current instance if it is produced during a time
 period containing $t$).

Similarly, a packet of previous instance for a
 query with a shorter period has a higher priority over  a packet of
 previous instance for a
 query with a longer period. Ties are broken by lexicographic order
 $\langle \mbox{current/previous, $\period_i$, ID} \rangle$.

$3)$ All packets of previous instances for the same query are scheduled
 on the first-in-first-out basis.

As proved in \cite{liu-layland-jacm73}, the rate monotonic method
 can achieve optimum performance for each packet to be transmitted
 before deadline,
 if each
 node has utilization (the utilization can be seen as
 the ratio of relay load to the fraction of time it is assigned to) of
 at most $n\cdot (2^{1/n}-1)$ where $n$ is the
 number of queries the node is involved.
Note that $n\le\qnum$. For large $n$, we obtain the utilization
 bound of $69\%$ means
 that as long as each node has utilization of less than $69\%$, all
 packets can make their deadlines.

\subsection{Sufficient Conditions for Schedulability}

In this section, we prove that the proposed algorithms for scheduling data collection queries
 are feasible.

\begin{lemma}\label{lem:col_intime}
The proposed algorithms can answer all
data collection queries if
\begin{eqnarray}\label{equ:col_suff}
\sum_{i}
 \frac{|\sources_i| \cdot \processing_i}{\period_i} \le
 \frac{0.69}{c_2({\cal M})\cdot c_3({\cal M})}
\end{eqnarray}
Here $c_2({\cal M})$ is the chromatic number for region coloring
 such that if we only select
 one node from each of the interference-aware regions with
 the same color to transmit, we can avoid interference under the interference model ${\cal M}$.
$c_3({\cal M}) >1$  is the maximum
size of CDS insider an interference-aware region plus one under the
 interference model ${\cal M}$.
 The value of $c_3({\cal M})$ is given as
\label{lem:c_3}
\begin{equation*}\label{equ:c_3}
c_3({\cal M})=
\begin{cases}
8\cdot(\rho+4)^2  & \text{under PrIM}\\
200  &  \text{under RTS/CTS}\\
200 & \text{under PhIM}.
\end{cases}
\end{equation*}
\end{lemma}

\begin{lemma}\label{lem:col_delay}
The proposed algorithms can answer all
queries within the deadlines,  if for each query, the end-to-end delay
 requirement  $\delay_i$ satisfies the inequality $\delay_i\ge c_2({\cal M})\cdot
T\cdot 2R$ where $R$ is the radius of
communication graph $G$.
\end{lemma}

Lemma~\ref{lem:col_intime} and \ref{lem:col_delay} imply
schedulability of the given set of queries. Thus, we propose a
sufficient condition for schedulability.
\begin{boldtheorem}\label{the:c_3}
Equation~(\ref{equ:col_suff}) is a sufficient condition for
schedulability of a set of data collection queries.
\end{boldtheorem}
We can illustrate by an example
 that the sufficient condition in Theorem~\ref{the:c_3} is almost tight.

 \begin{figure}[h]
\begin{center}
\scalebox{0.25}{\input{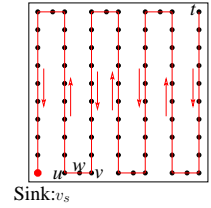}}
\caption{An example for node placement in an interference-region.}
\label{fig:col bound}
\end{center}
\end{figure}

In Figure~\ref{fig:col bound}, node $v_s\in V$ is the sink. There
are $\sqrt{ c_3({\cal M})}$ vertical evenly spaced lines with
 distance $d_h =(1+\epsilon)$ between consecutive lines
 (\eg, the distance between $u$ and $v$ is $1+\epsilon$).
 Here we simply assume $\sqrt{ c_3({\cal M})}$ is an integer.
Additionally, ${\sqrt{ c_3({\cal M})}}-1$ nodes, like $w$ between $u$ and $v$, act
as
 bridges to keep the network connectivity.
Clearly, there are $\sqrt{ c_3({\cal M})}=O(\rho^2)$ nodes deployed in the
 interference-aware region, and the size of CDS in this region is
 $c_3({\cal M})-1$.
The residual network (all nodes outside of the region)
 is connected to sink $v_s$ only through node $t$.
We assume all sources nodes ($\sources_j$ for the $j$-th query)
 are located in the residual network.
To collect data to the sink $v_s$,
 we should strictly follow the red path.
It is easy to verify that
 the relay load of the interference-aware region is  $c_3({\cal M})
 \cdot \sum_i
 \frac{|\sources_i| \cdot\processing_i}{\period_i}$ (the initial load is zero).
Thus a necessary condition for schedulability for the example
network
 is $\sum_{i} \frac{|\sources_i| \cdot\processing_i}{\period_i}\le
 \frac{c_1({\cal M})}{c_3({\cal M})}$.
We can verify that the sufficient condition in
 Theorem~\ref{lem:col_intime}
tightly match this necessary condition
 by a factor of at most $c_1({\cal M})\cdot c_2({\cal M})$
which is independent of $c_3({\cal M})$.

\begin{figure*}[t]
\centering
\begin{tabular}{cc}
\epsfxsize=3.3in\epsfysize=2.7in\epsfbox{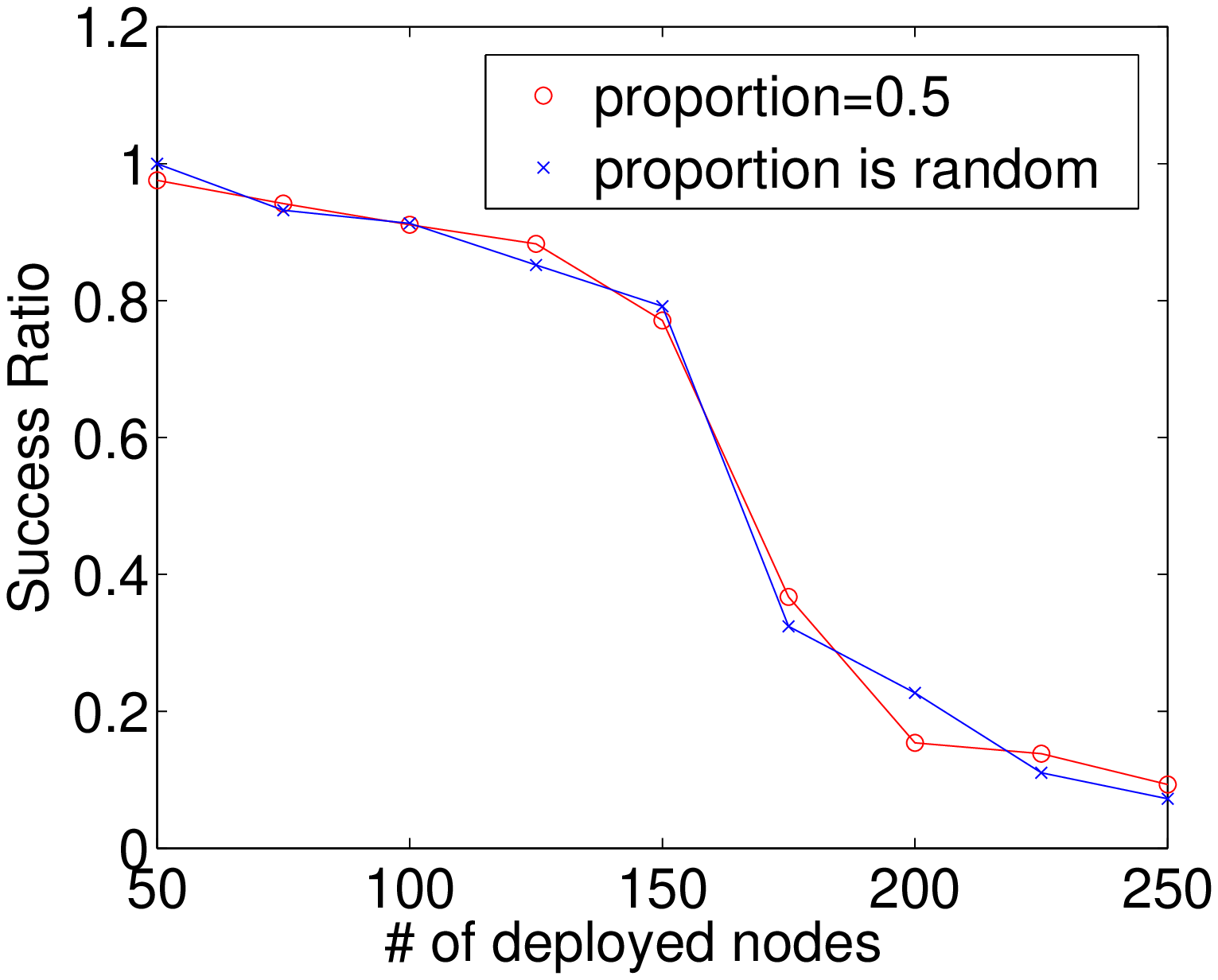} & \epsfxsize=3.3in\epsfysize=2.7in\epsfbox{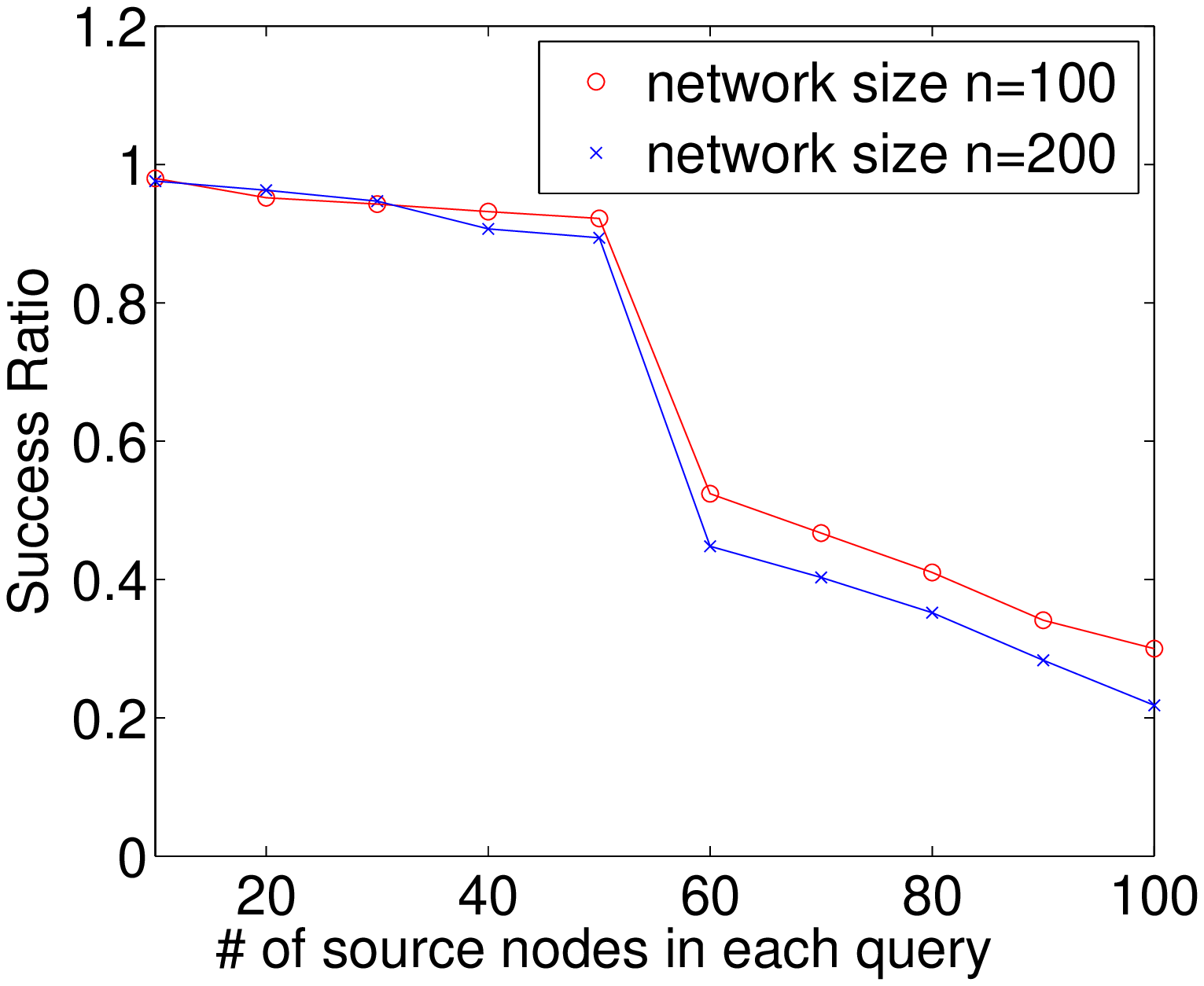} \\
(a) Network size increases & (b) \# of sources increases
\end{tabular}
\caption{Performances of data collection algorithm. In (a), 'proportion' denotes the ratio of the number of
 source nodes over the total number of nodes.}
\label{fig:collection_result}
\end{figure*}

\section{Drop Overloaded Queries}\label{sec:weighted}

\def\knap{\texttt{KS}}

\begin{algorithm}[b]
\caption{Maximum Weighted Query Selection}
\begin{algorithmic}[1]
\STATE $A_{[1]}:=\{\arg\max_{i; \frac{|\sources_i| \cdot
\processing_i}{\period_i}\leq 1}\{\weight_i\}$\}; \STATE $A_{[2]}:=
\mbox{ the solution returned by }\knap(\frac{0.69}{c_2({\cal
M})\cdot c_3({\cal M})})$; \STATE $A:=\arg\max_{A\in
\{A_{[1]},A_{[2]}\}} \{\weight(A)\}$;
\end{algorithmic}
\label{alg:weight}
\end{algorithm}

In this section, we study scheduling for
 an overloaded sensor network when not all arriving
 queries can be scheduled.
Let us focus on the data collection queries: given a set of data
collection queries $\queryset$,
 assume the $i$-th query  is associated with a weight $\weight_i$.
The objective is to select and schedule a subset of queries
$S\subseteq \queryset$  to maximize the overall weight of the
scheduled queries.

We reduce our problem to a \emph{0-1 knapsack problem} as follows:
given $\qnum$ items,
 the $i$-th query can be considered as an item of size
$\frac{|\sources_i| \cdot \processing_i}{\period_i}$ and weight
$\weight_i$. The objective is to select a subset of items with total
size at most $C$ such that the weighted sum of all selected items is
maximized. Here $C$ is called the bag size. We will denote the
$0$-$1$ knapsack problem with bag size $C$ by
 $\knap(C)$ for brevity.

Then, our algorithm consists of two phases:

\textbf{Phase I:} we enumerate each single query whose load
$\frac{|\sources_i| \cdot \processing_i}{\period_i}$ is no larger
than $1$ and select the one with the maximum weight as the first
candidate solution;

\textbf{Phase II:} we use the solution for
$\knap(\frac{0.69}{c_2({\cal M})\cdot c_3({\cal M})})$ as the second
candidate solution.

The final solution can be obtained by choosing the one
 with larger weight among these two candidate solutions.
Please refer to Algorithm \ref{alg:weight} for details. Note that we
can design a joint routing and scheduling protocol to satisfy a set
of data collection queries $\queryset$ under an interference model
$\cal M$, if $ \sum_{i}
 \frac{|\sources_i| \cdot \processing_i}{\period_i} \le
 \frac{0.69}{c_2({\cal M})\cdot c_3({\cal M})}$.
Therefore, it is easy to verify the correctness of our solution.

The challenge here is to derive an approximation bound on this
solution. Recall that for any set of schedulable queries, we must
have $\sum_{i}
 \frac{|\sources_i| \cdot \processing_i}{\period_i}  \le 1$,
which implies that the optimal solution for our problem is no larger
than the optimal solution of $\knap(1)$. Let $OPT_{\knap(1)}$ denote
the optimal solution of $\knap(1)$. The following lemma shows that
the selected queries have weight at least a constant fraction of the
weight of $OPT_{\knap(1)}$.

\begin{lemma}\label{lem:weighted}
Let $\weight(A)$ denote the weight of the queries selected by
Algorithm \ref{alg:weight}, and $d=\frac{0.69}{c_2({\cal M})\cdot
c_3({\cal M})}$, we have
$\frac{d}{2}\cdot\weight(OPT_{\knap(1)})\leq \weight(A)$.
\end{lemma}
Together with the fact that the optimum solution of our problem is
no larger than $\weight(OPT_{\knap(1)})$,
Theorem~\ref{the:weighted_appr} immediately follows.
\begin{boldtheorem}\label{the:weighted_appr}
Algorithm \ref{alg:weight} is $d/2$-approximation for the maximum
weighted query selection problem, where
 $d=\frac{0.69}{c_2({\cal M})\cdot c_3({\cal M})}$.
\end{boldtheorem}

In the previous discussions, we assumed that we will drop some
queries when we cannot answer all queries in time. In practice, it
may be
 possible to partially satisfy all queries, by carefully
 dropping some packets from some query flows (once every certain period), or
 dropping some packets from some data-source nodes.
  Dropping packets (temporally or spatially)
 is feasible for some applications because of the possible
 (temporal and/or spatial) correlation among data sensed by different
 sensors.
Our algorithm can also be extended to deal with this case and
details
 are omitted due to space limitations.

\section{Simulation results}
\label{sec:simulation}

We randomly deploy a set of nodes $\{v_1,\cdots, v_n \}$ with
transmission range $50$
 in an area of size $400 \times 400$ (note that we always keep
 connectivity of the networks).
For any pair of nodes $v_i$ and $v_j$, there is a feasible link
 if $|v_iv_j|\le 50$.
In addition, each link $(v_1,v_2)$ is associate with a quality
variable $q_{v_1v_2}$. Here, the value of $q_{v_1v_2}$
 is proportion to $|v_iv_j|$.

The main flow of our evaluation system is as follows: The sink node
will generate up to $20$ data collection
 queries and broadcast it to the network one by one.
The broadcast procedure will not stop until all source nodes in the
 receive  the $i$-th query correctly.
Secondly, the sink node initiates
 to construct
 routing trees (based on the CDS) which cover all source nodes (may need non-source nodes
   to relay).
After a certain starting time, each source node will read the corresponding data
  repeatedly 
  and transmit via routing trees.
The sink node will continue to analyze all received data packets
 for each period of each query.
When all currently existing queries are satisfied, the sink node
 will release next query up to $20$ queries totally.
The algorithm will
terminate when none of existing queries is satisfied,

We now evaluate the performance of our algorithms in different
scenarios. In the first scenario,
 we vary the network size from $50$ to $250$ with step $25$.
For each query, we pick source nodes randomly or always choose a set
of
 source nodes with half of the network size.
Figure \ref{fig:collection_result}(a) shows the results when we
either randomly pick the number
 of source nodes for each query or always randomly pick half of the nodes as
 source nodes.
The success ratio is equal to the number of successful rounds
divided by the total rounds.

When the network size increases over $150$, the success ratio will
quickly
 drop from around $0.8$ to $0.35$.
This is mainly caused by capacity bounds of CDS. The new packets
from newly increased nodes (hence newly increased source nodes) lead
CDS
 saturated such that many packets are dropped due to the buffer limit.

In the second scenario, we fix the network size and increase the
number of source nodes
 in each query from $10$ to $100$ with step $10$.
The figure \ref{fig:collection_result} (b) shows the success ratio
when the network size is
 $100$ and $200$ respectively.
As we can see, when the number of source nodes is small (less than
$50$),
 most queries are satisfied.
When the number of source nodes is larger than $50$, the performance
dropped quickly. In addition, there is no big difference when
network sizes ($100$ and $200$ respectively) are
 different.

\section{Conclusions}
\label{sec:conclusion}

We proposed joint design of a family of routing and packet
scheduling schemes under different interference models. Most
importantly, we theoretically proved that our algorithm can achieve
constant approximation in terms of schedulability. We also studied
the overloaded case where not all queries can be scheduled by
proposing an efficient method for carefully selecting a subset of
queries that maximizes the overall weight of the scheduled queries.
In this case, we theoretically proved that our proposed scheme
 can achieve constant approximation.

\section*{Acknowledgment}
The research of X. Xu and M. Song is supported in part by NSF CAREER Award CNS-1248092.
The research of M. Song is also supported by NSF IPA Independent Research and Development (IR/D) Program.
The research of Xiang-Yang Li
  is partially supported by NSF CNS-0832120, NSF CNS-1035894, NSF ECCS-1247944,
  National Natural Science Foundation of China under Grant No. 61170216, No. 61228202, China 973 Program under Grant No.2011CB302705.
However, any opinion, finding, and conclusions or recommendations expressed in this
material, are those of the author and do not necessarily reflect the views of
the funding agencies (NSF and NSFC).

\end{document}